\documentclass{codee}
\usepackage{graphicx}
\usepackage[numbers]{natbib}
\usepackage{hyperref}
\newtheorem{theorem}{Theorem}[section]

\theoremstyle{definition}

\newtheorem{problem}[theorem]{Problem}

\theoremstyle{remark}

\numberwithin{equation}{section}

\title{ODE models of wealth concentration and taxation} 
\shorttitle{Model Document}
\author{Bruce Boghosian and Christoph B\"orgers}
\affiliation{Department of Mathematics, Tufts University, Medford, Massachusetts}
\shortauthor{One, Two and Three}
\date{July 8, 2016}
\dedicatory{}
\keywords{wealth inequality, oligarchy, transcritical bifurcation}
\submissiondate{???}
\publicationdate{???}
\articlenumber{???}

\begin{document}
\maketitle

% Please type in a short abstract here. Please try to avoid using any mathematical symbols
% as the abstract will also go on the CODEE Digital Library.
\begin{abstract}
We refer to an individual holding a non-negligible fraction of the country's total wealth as an {\em oligarch.}
We explain how a 
 model due to Boghosian {\em et al.}\   \cite{boghosian_et_al} can be used to explore the effects of taxation on the emergence of oligarchs. The model suggests that
oligarchs will emerge when  {\em wealth} taxation is below a certain threshold, not when it is above the threshold. The underlying
mechanism is a transcritical bifurcation. The model also suggests that taxation of {\em income and capital
gains} alone cannot prevent the emergence of oligarchs. We suggest several opportunities for
students to explore modifications
of the model. 
\end{abstract}

% Your paper really starts here. Use \section and \subsection to
% create section headings.

\section{Introduction}

The concentration of wealth in the hands of a small number of individuals 
is considerable in many countries around the world, including the United States.
In 2020, according to 
ref.\ \citep{williamson}, the 400 wealthiest Americans held 3.5\% of the country's total wealth. 
(The same article also  estimated all Black 
households combined to hold a total of 3\% of all U.S.\ wealth.) We present here a
 model extracted from 
Boghosian {\em et al.}\ \cite{boghosian_et_al}, 
which sheds light on the effects of  taxation on wealth concentration. 

Individuals who hold a significant fraction of society's total wealth, and therefore considerable political power \cite{krugman_2020}, will be called {\em oligarchs} here.
The model suggests the existence of a minimal rate of wealth taxation below which oligarchs may emerge.
The underlying mathematical mechanism  is a {\em transcritical bifurcation} \cite{strogatz_book}. 

We present and analyze the model in Section 2, discuss some variations in Section 3, and suggest problems
for students to think about in Section 4. We very briefly sketch in Section 5 what is captured by the full model 
and analysis
in \cite{boghosian_et_al}, but is not reproduced in our discussion here.

\vfill
\pagebreak

\section{Oligarchy and a flat wealth tax} 

We start by assuming that there are oligarchs holding a fraction $w \in (0,1)$ of society's wealth. 
We assume that in a short time $dt$, the oligarchs, by virtue of the  power that their wealth
gives them, acquire  a small fraction $\lambda dt$ of the wealth not yet in their hands:
\begin{equation}
\label{eq:exponential}
\frac{dw}{dt}  = \lambda (1-w).
\end{equation}
As $w$ increases, acquiring wealth becomes easier for the oligarchs: They can hire more lobbyists and better laywers, they can afford riskier but more profitable investments, and so on. It is therefore reasonable
to assume that $\lambda$  is proportional to $w$, so 
\begin{equation}
\label{eq:rw} 
\lambda =g w ~~\mbox{for some  constant $g >0$}.
\end{equation}
You might think of $g$ as measuring the rate at which the oligarchs grab what's not
theirs yet; hence $g$, the first letter in ``grab". 
We thereby arrive at the logistic growth equation
\begin{equation}
\label{eq:logistic_equation}
\frac{dw}{dt} = g w (1-w). 
\end{equation}

\begin{figure}[h!]
\begin{center}
\includegraphics[scale=0.6]{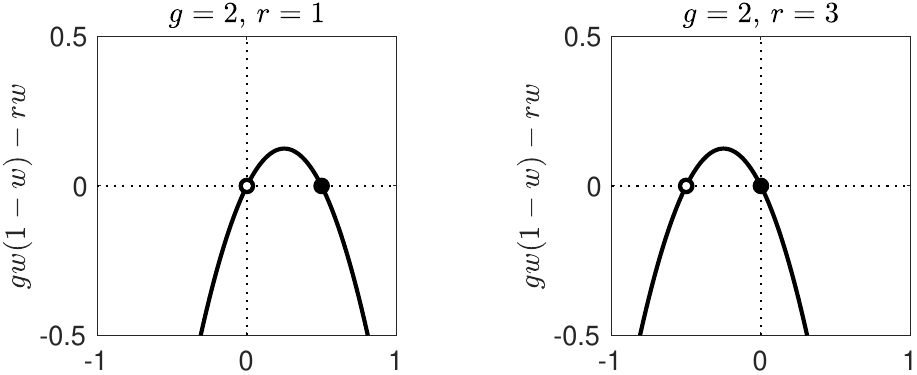}
\caption{The right-hand side of equation 
 \ref{eq:logistic_with_death_rate} for $r<g$ (left) and for 
$r > g$ (right).}
\label{fig:RHS} 
\end{center}
\end{figure}

Suppose now that the government taxes the oligarchs, taking away a small fraction $r  dt$ of their wealth
in time $dt$, where $r>0$ is another parameter.  We use $r$  because it is the first
letter in ``rate" (of taxation). This is 
called a {\em flat} tax because $r$ does not dependent on $w$. 
The equation now becomes 
\begin{equation}
\label{eq:logistic_with_death_rate}
\frac{dw}{dt} = g w (1-w) - r w.
\end{equation}
Equation \ref{eq:logistic_with_death_rate} has two fixed points, obtained by setting the right-hand side equal to zero and solving for $w$: 
\begin{equation}
\label{eq:fixed_points}
w=0 ~~~\mbox{and}~ ~~ w = 1 - \frac{r}{g}. 
\end{equation}
This is illustrated in Figure \ref{fig:RHS}.

\begin{figure}[h!]
\begin{center}
\includegraphics[scale=0.25]{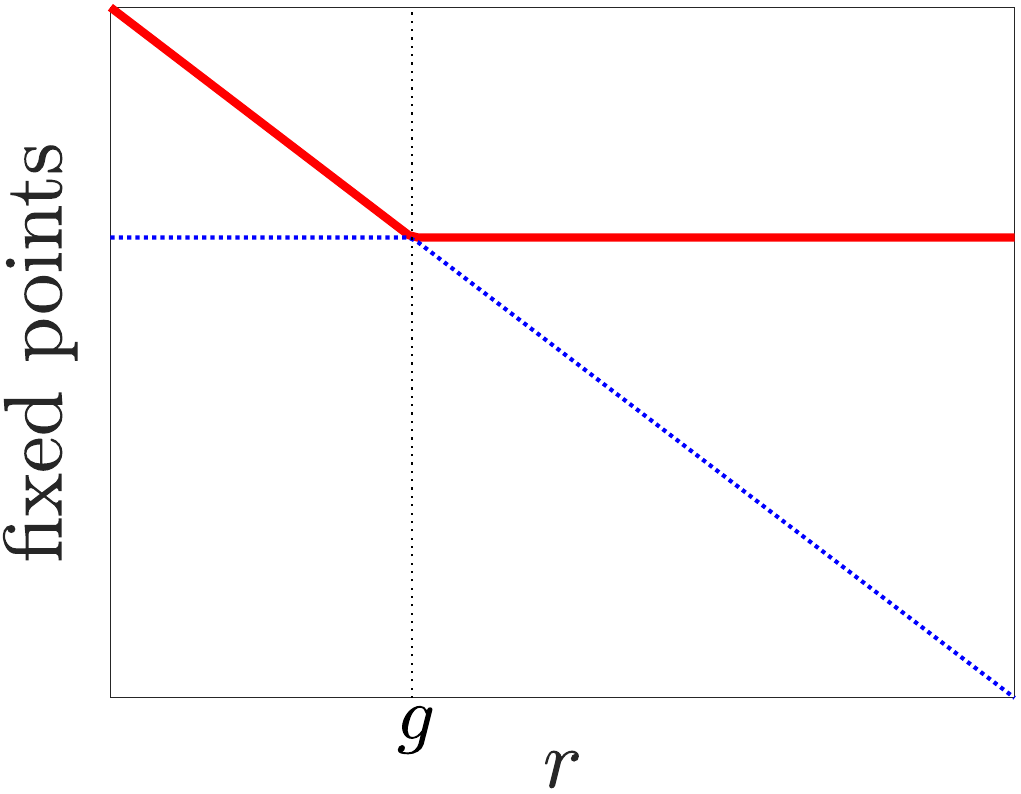}
\caption{The fixed points as a function of $r$ for a fixed $g$. The attracting fixed point is indicated in red and
bold.} 
\label{fig:TRANSCRITICAL} 
\end{center}
\end{figure}

For $r>g$, the fixed point $w=0$ is attracting, and
$1 - \frac{r}{g} $ is negative and repelling. In keeping with common convention, this is indicated 
by open circles and closed circles in Figure \ref{fig:RHS}. 
For $r<g$, the fixed point
$w=1 - \frac{r}{g}$ is positive and attracting, and $0$ is repelling. The interpretation is that
for $r>g$ there will eventually be no oligarchs.  For $r<g$, there
will be oligarchs (the attracting fixed point is positive). 
We have here an  example of a {\em transcritical bifurcation} \cite{strogatz_book}. The two fixed points ``collide" 
as $r$ moves through $g$, and ``exchange their stability properties"; see Figure \ref{fig:TRANSCRITICAL}.

\begin{figure}[h!]
\begin{center}
\includegraphics[scale=0.25]{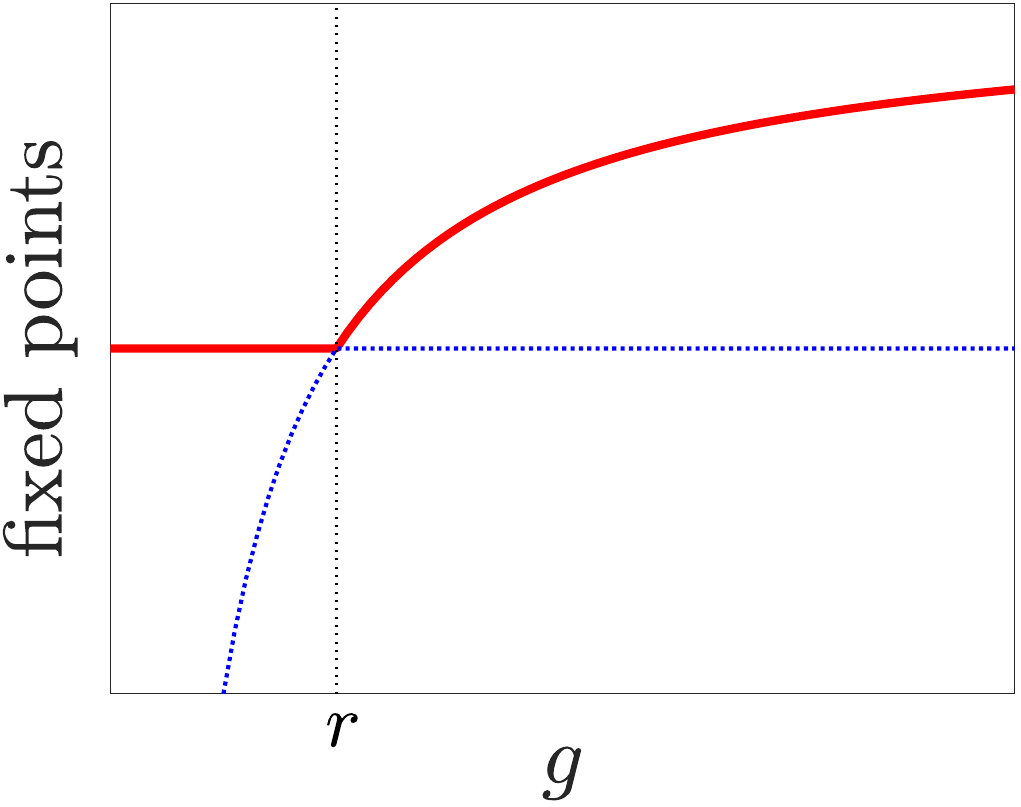}
\caption{The fixed points as a function of $g$ for a fixed $r$. The attracting fixed point is indicated in red and
bold.} 
\label{fig:TRANSCRITICAL_2} 
\end{center}
\end{figure}

The 
transcritical bifurcation that occurs as $r$ rises above $g$ can also be achieved by reducing $g$ below $r$, that is, by reducing
the wealth-acquired advantage --- perhaps 
through public policy measures such as regulations on lobbying \cite{lobbying}. See Figure \ref{fig:TRANSCRITICAL_2}, and compare with  \cite[Figure 1]{boghosian_et_al}.

In summary, the model suggests that taxing wealth sufficiently heavily can prevent oligarchy.

\vfill 
\pagebreak

\section{Variations} 

\subsection{Flat income tax} 

To model the taxation of 
{\em income} (or capital gains),  assume the government takes  away a fixed fraction $\theta \in (0,1)$ of the wealth gained by
the oligarchs in time $dt$:
\begin{equation}
\label{eq:income_tax}
\frac{dw}{dt} = (1-\theta) g w (1-w). 
\end{equation}
If $0 < \theta < 1$ and $g>0$, equation  \ref{eq:income_tax} has the attracting fixed point $w=1$ and the
repelling fixed point $w=0$. The model therefore suggests that 
income taxation alone does not prevent (total) oligarchy in the long run. 

\subsection{Progressive wealth tax} 

\begin{figure}[h!]
\begin{center}
\includegraphics[scale=0.6]{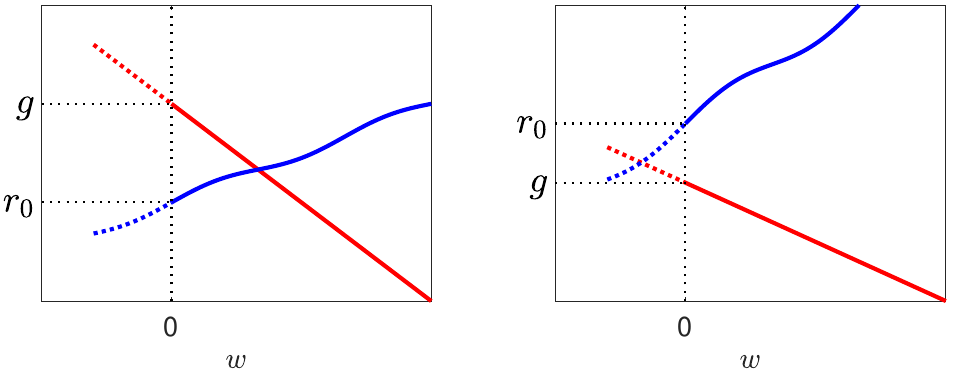}
\caption{The left (red) and right (blue) sides of equation \ref{eq:progressive_fixed_points}, for $r_0<g$
and $r_0>g$.}
\label{fig:TWO_SIDES} 
\end{center}
\end{figure}

Progressive wealth taxation would mean that $r$ itself grows with $w$. Assume $r=r_0 P(w)dt $, where $r_0$
is a fixed positive constant, and $P$ is a continuous,\footnote{We should assume that $P$ is {\em Lipschitz} continuous to 
remove any doubts regarding the existence and uniqueness of solutions of equation \ref{eq:progressive_wealth_tax}, but such technical points are not our focus here.}  monotonically increasing function of $w \in [0,1]$ with $P(0)=1$. 
The equation now becomes 
\begin{equation}
\label{eq:progressive_wealth_tax}
\frac{d w}{dt} = g w (1- w) -  r_0 P(w) w.
\end{equation}
Fixed points are obtained by setting the right-hand side equal to $0$. One fixed point is always $w=0$, corresponding to the absence 
of oligarchs. Other fixed points, if there are any, must solve 
\begin{equation}
\label{eq:progressive_fixed_points}
g (1-w) = r_0 P(w).
\end{equation}
To understand the solutions of equation \ref{eq:progressive_fixed_points}  we plot the left-hand side as a function 
of $w \in [0,1]$, and, in the same plot, the right-hand side as well. Solutions of equation \ref{eq:progressive_fixed_points} are the abscissae of points where the two graphs meet. Examples are shown in Figure \ref{fig:TWO_SIDES}. 
If $r_0 < g$, there exists a positive fixed point and it is attracting, while the fixed point $0$ is repelling then. 
If $r_0>g$, there exists a negative fixed point, but no positive one, and it is repelling, while the fixed point
$0$ is attracting.

The conclusion is that a progressive wealth tax at rate $r_0 P(w)$ yields the same threshold as a flat wealth tax 
at rate $r_0$: 
Oligarchs emerge
for $r_0<g$, and not for $r_0>g$. However, for $r_0<g$, the positive fixed point 
will be smaller with a progressive wealth tax than it would be with a flat wealth tax at rate $r_0$, so the 
oligarchs will hold less of society's wealth;  see Problem \ref{problem:progressive}.

\vfill
\pagebreak

\section{Suggested problems} 

\begin{problem} {\bf Modified model of the wealth-acquired advantage.} Equation \ref{eq:rw} seems a bit
 arbitrary. While 
it seems plausible that $\lambda$ should be an increasing function of $w$, why should it be proportional to $w$? 
What happens if $\lambda = g w^2$ for instance? Then equation \ref{eq:logistic_with_death_rate} is 
replaced by 
$$
\frac{dw}{dt} = g w^2 (1-w) - r w.
$$
Determine the fixed points of this equation, study the bifurcation that occurs as $r$ is raised or $g$ is lowered, and interpret the results in terms of the emergence or absence of oligarchs.
Try other modifications that make sense to you as well. 
\end{problem} 

\begin{problem} {\bf Progressive income tax.} Suppose that the fraction of the oligarchs' gains taken 
away by the government is not a fixed $\theta \in (0,1)$, but rather 
$$
Q \left( g w (1-w) \right)
$$
where $Q=Q(x)$ is a monotonically increasing function of $x \geq 0$ with $Q(0) \geq 0$. The greater the oligarchs' gain
in a unit of time, the greater  a percentage of their gains will be taken away from them.  
You may want to assume that the government will never take away as much as 
100\% of the oligarchs' gains.\footnote{The Swedish children's book author Astrid Lindgren complained in 1976
that her {marginal} tax rate had risen to 102\%. She wrote a satirical story about it, published in the 
Swedish tabloid {\em Expressen} \cite{Lindgren}. It should be 
said that there is a distinction between the {\em marginal} tax rate --- the rate at which the last dollar of income is taxed --- and the {\em effective}  one.  Lindgren, by the way, supported the Swedish Social Democrats, 
who had introduced  those steeply progressive income tax rates, throughout  her life, even after 1976.}, so 
$Q(x)<1$ for all $x$.
Now the equation is 
$$
\frac{dw}{dt} =( 1 - Q(g w (1-w))) g w (1-w).
$$
Will oligarchs be able to persist? 
\end{problem}

\begin{problem} {\bf Progressive wealth tax re-visited.} Assume a 
progressive wealth tax with $r_0 < g$.  Explain why the positive fixed point of equation \ref{eq:logistic_with_death_rate} is closer to zero
than it would be with a flat wealth tax at rate $r_0$.
\label{problem:progressive} 
\end{problem} 

\begin{problem} {\bf Regressive wealth tax.} Suppose that as $w$ increases, the oligarchs, by virtue of their increasing
political power, manage
to reduce the wealth tax rate. So the equation is \ref{eq:progressive_wealth_tax}, but now
$P$ is a {\em decreasing} function of $w$ with $P(0) = 1$ and $P(w) \geq 0$. Analyze what happens. (It will 
depend on what you assume about the function $P$.) 
\end{problem}

\section{What the full model is about} 

We mentioned that the model suggested here is extracted from 
a model due to Boghosian {\em et al.} \cite{boghosian_et_al}. What would you learn by reading 
\cite{boghosian_et_al} that you cannot learn here?
The essential answer is that you would learn about the {\em distribution} of wealth, not merely 
about the fraction of wealth held by the oligarchs. 
As a result, the model of \cite{boghosian_et_al} predicts things like  \href{https://www.youtube.com/watch?v=y8y-gaNbe4U}{the Lorenz curve and the Gini coefficient.} 
Our equation \ref{eq:logistic_with_death_rate}, supported
here by plausibility arguments only, is {\em derived} in \cite{boghosian_et_al} (see equation 17 in \cite{boghosian_et_al}) from a model of the time evolution of the wealth distribution.

\vfill
\pagebreak

\begin{center}
{\Large Sketches of solutions for the problems} 
\end{center}

The  problems are intended to inspire the reader to investigate on their own, and the solution
sketches given here are not meant to undermine that, but to assist a reader who would like help getting started.

\vskip 10pt
\noindent
{\bf 4.1. Modified model of the wealth-acquired advantage.}  For the modified equation, $w=0$ is still 
a fixed point. Since the linearization of $g w^2 (1-w) - r w$ around $w=0$ is $-r w$, 
the fixed point $w=0$ is always attracting now. Other fixed points must satisfy 
$$
w (1-w) = \frac{r}{g}.
$$
Plot $w (1-w)$ as a function of $w$ to see that there are no solutions of this equation when 
$r > \frac{g}{4}$, and there are two solutions, both strictly between $0$ and $1$, when 
$r < \frac{g}{4}$. The larger of the two positive fixed points is then attracting, and the 
smaller is repelling. If $w$ is perturbed from $0$
slightly, it returns to the attracting fixed point $0$, but if $w$ is raised beyond the smaller positive fixed point,
which acts as a threshold, it converges to the larger one. This is an example of 
 an {\em excitable} system \cite{strogatz_book}.
The bifurcation that annihilates the two 
positive fixed points as $r$ rises above $\frac{g}{4} $ is a {\em saddle-node bifurcation} 
\cite{strogatz_book}.

\vskip 10pt
\noindent
Now think about other possible relations between $\lambda$ and $w$.

\vskip 10pt
\noindent
{\bf 4.2.  Progressive income tax.} If $Q(x)<1$ for all $x$, the only two fixed points are still $w=0$ and $w=1$, and $w=0$ is repelling,
$w=1$ is attracting. But what if we allow $Q(x)>1$? 

\vskip 10pt
\noindent
{\bf 4.3. Progressive wealth tax re-visited.} Ask yourself what Figure 4 would look like if the wealth tax were flat.

\vskip 10pt
\noindent
{\bf 4.4\ Regressive wealth tax.} The simplest example would be 
$P(w) = 1-w$. Equation \ref{eq:progressive_wealth_tax} then becomes 
$$
\frac{dw}{dt} =  (g- r_0) w (1-w).
$$
This equation has fixed points at $w=0$ and $w=1$. For $ r_0 < g $, the fixed point at $0$ is repelling,
and that at 1 is attracting. For $r_0 > g$, the fixed point at $0$ is attracting, and that at 1 is repelling.
So even the regressive wealth tax with $P(w) = 1-w$ still leads to the conclusion that oligarchy emerges if 
$r_0 < g$, but not if $r_0 > g$.

\vskip 5pt
\noindent
If you were to make plots like those in Figures \ref{fig:TRANSCRITICAL} and \ref{fig:TRANSCRITICAL_2} 
(with $r_0$ replacing $r$), 
what would they look like now? Is it still a transcritical bifurcation?

\vskip 5pt
\noindent
Next think about other examples, for instance $P(w) = 1-w^2$; that would mean 
that the oligarchs become really effective at fighting the wealth tax only when they have acquired enough
wealth.

\end{document}